\documentstyle[12pt]{article}

\def\doit#1#2{\ifcase#1\or#2\fi}

\skewchar\fivmi='177
\skewchar\sixmi='177
\skewchar\sevmi='177
\skewchar\egtmi='177
\skewchar\ninmi='177
\skewchar\tenmi='177
\skewchar\elvmi='177
\skewchar\twlmi='177
\skewchar\frtnmi='177
\skewchar\svtnmi='177
\skewchar\twtymi='177
\def\@magscale#1{ scaled \magstep #1}

\def\framingfonts#1{
\doit{#1}{\font\twfvmi  = ammi10   \@magscale5 
\skewchar\twfvmi='177
\skewchar\fivsy='60
\skewchar\sixsy='60
\skewchar\sevsy='60
\skewchar\egtsy='60
\skewchar\ninsy='60
\skewchar\tensy='60
\skewchar\elvsy='60
\skewchar\twlsy='60
\skewchar\frtnsy='60
\skewchar\svtnsy='60
\skewchar\twtysy='60
\font\twfvsy  = amsy10   \@magscale5 
\skewchar\twfvsy='60
\font\go=font018			
\font\sc=font005			
\def\Go#1{{\hbox{\go #1}}}	
\def\Sc#1{{\hbox{\sc #1}}}	
\def\Sf#1{{\hbox{\sf #1}}}	
\font\oo=circlew10	      
\font\ooo=circle10			
\font\ro=manfnt				
\def\kcl{{\hbox{\ro 6}}}		
\def\kcr{{\hbox{\ro 7}}}		
\def\ktl{{\hbox{\ro \char'134}}}	
\def\ktr{{\hbox{\ro \char'135}}}	
\def\kbl{{\hbox{\ro \char'136}}}	
\def\kbr{{\hbox{\ro \char'137}}}	
}}

\catcode`@=11
\def\un#1{\relax\ifmmode\@@underline#1\else
	$\@@underline{\hbox{#1}}$\relax\fi}
\catcode`@=12

\let\du=\d			
\let\um=\H			

\def\a{\alpha}
\def\b{\beta}

\def\d{\delta}
\def\e{\epsilon}

\def\g{\gamma}

\def\l{\lambda}
\def\m{\mu}

\def\r{\rho}
\def\s{\sigma}
\def\t{\tau}

\def\G{\Gamma}

\def\L{\Lambda}

\def\plpl{{\raise-2pt\hbox{$\raise3pt\hbox{$_+$}\hskip-7.0pt\raise0.0pt
\hbox{$^+$}\hskip 0.01pt$}}}
\def\mimi{{\raise-2pt\hbox{$\raise3pt\hbox{$_-$}\hskip-7.0pt\raise0.0pt
\hbox{$^-$}\hskip 0.01pt$}}}

\def\bo{{\raise.15ex\hbox{\large$\Box$}}}		
\def\pr{\prod}						
\def\TH{{\raise.2ex\hbox{$\displaystyle \bigodot$}\mskip-4.7mu \llap H \;}}
\def\face{{\raise.2ex\hbox{$\displaystyle \bigodot$}\mskip-2.2mu \llap {$\ddot
	\smile$}}}					

\def\sp#1{{}^{#1}}				
\def\Tilde#1{{\widetilde{#1}}\hskip 0.03in}			
\def\Hat#1{\widehat{#1}}			
\def\Bar#1{\overline{#1}}			
\def\leftrightarrowfill{$\mathsurround=0pt \mathord\leftarrow \mkern-6mu
	\cleaders\hbox{$\mkern-2mu \mathord- \mkern-2mu$}\hfill
	\mkern-6mu \mathord\rightarrow$}
\def\dvec#1{\vbox{\ialign{##\crcr
	\leftrightarrowfill\crcr\noalign{\kern-1pt\nointerlineskip}
	$\hfil\displaystyle{#1}\hfil$\crcr}}}		
\def\dt#1{{\buildrel {\hbox{\LARGE .}} \over {#1}}}	

\def\frac#1#2{{\textstyle{#1\over\vphantom2\smash{\raise.20ex
	\hbox{$\scriptstyle{#2}$}}}}}			
\def\sfrac#1#2{{\vphantom1\smash{\lower.5ex\hbox{\small$#1$}}\over
	\vphantom1\smash{\raise.4ex\hbox{\small$#2$}}}}	
\def\bfrac#1#2{{\vphantom1\smash{\lower.5ex\hbox{$#1$}}\over
	\vphantom1\smash{\raise.3ex\hbox{$#2$}}}}	
\def\afrac#1#2{{\vphantom1\smash{\lower.5ex\hbox{$#1$}}\over#2}}    

\newskip\humongous \humongous=0pt plus 1000pt minus 1000pt
\def\caja{\mathsurround=0pt}
\def\eqalign#1{\,\vcenter{\openup2\jot \caja
	\ialign{\strut \hfil$\displaystyle{##}$&$
	\displaystyle{{}##}$\hfil\crcr#1\crcr}}\,}
\newif\ifdtup
\def\panorama{\global\dtuptrue \openup2\jot \caja
	\everycr{\noalign{\ifdtup \global\dtupfalse
	\vskip-\lineskiplimit \vskip\normallineskiplimit
	\else \penalty\interdisplaylinepenalty \fi}}}
\def\li#1{\panorama \tabskip=\humongous				
	\halign to\displaywidth{\hfil$\displaystyle{##}$
	\tabskip=0pt&$\displaystyle{{}##}$\hfil
	\tabskip=\humongous&\llap{$##$}\tabskip=0pt
	\crcr#1\crcr}}

\def\ref#1{$\sp{#1)}$}

\parskip=\medskipamount			
\thicklines			    

\def\oldheadpic{				
	\setlength{\unitlength}{.4mm}
	\thinlines
	\par
	\begin{picture}(349,16)
	\put(325,16){\line(1,0){4}}
	\put(330,16){\line(1,0){4}}
	\put(340,16){\line(1,0){4}}
	\put(335,0){\line(1,0){4}}
	\put(340,0){\line(1,0){4}}
	\put(345,0){\line(1,0){4}}
	\put(329,0){\line(0,1){16}}
	\put(330,0){\line(0,1){16}}
	\put(339,0){\line(0,1){16}}
	\put(340,0){\line(0,1){16}}
	\put(344,0){\line(0,1){16}}
	\put(345,0){\line(0,1){16}}
	\put(329,16){\oval(8,32)[bl]}
	\put(330,16){\oval(8,32)[br]}
	\put(339,0){\oval(8,32)[tl]}
	\put(345,0){\oval(8,32)[tr]}
	\end{picture}
	\par
	\thicklines
	\vskip.2in}
\def\oldtitle#1#2#3#4{\oldheadpic\begin{center}\vglue.5in{\large\bf #1}\\[.6in]
	{#2}\\[.1in] {\it Department of Physics and Astronomy}\\
	{\it University of Maryland, College Park, MD 20742}\\[.6in]
	Physics Publication \#{#3}\\ {#4}\\[1.5in] {\bf Abstract}\\[.1in]
	\end{center} \begin{quotation}}			
\def\oldTitle#1#2#3#4#5#6#7{\oldheadpic\begin{center} \vglue .4in
	{\large\bf #1}\\[.4in]
	{#2}\\[.1in] {\it Department of Physics and Astronomy}\\
	{\it University of Maryland, College Park, MD 20742}\\[.1in]
	{#3}\\[.1in] {\it {#4}}\\ {\it {#5}}\\[.4in]
	Physics Publication \#{#6}\\ {#7}\\[.5in] {\bf Abstract}\\[.1in]
	\end{center} \begin{quotation}}			
\def\border{						
	\setlength{\unitlength}{1mm}
	\newcount\xco
	\newcount\yco
	\xco=-24
	\yco=12
	\begin{picture}(140,0)
	\put(\xco,\yco){$\ktl$}
	\advance\yco by-1
	{\loop
	\put(\xco,\yco){$\kcl$}
	\advance\yco by-2
	\ifnum\yco>-240
	\repeat
	\put(\xco,\yco){$\kbl$}}
	\xco=158
	\yco=12
	\put(\xco,\yco){$\ktr$}
	\advance\yco by-1
	{\loop
	\put(\xco,\yco){$\kcr$}
	\advance\yco by-2
	\ifnum\yco>-240
	\repeat
	\put(\xco,\yco){$\kbr$}}
        \put(-20,11){\tiny University of Maryland Elementary Particle
Physics University of Maryland Elementary Particle Physics University of
Maryland Elementary Particle Physics}
	\put(-20,-241.5){\tiny University of Maryland Elementary
Particle Physics University of Maryland Elementary Particle Physics
University of Maryland Elementary Particle Physics}
	\end{picture}
	\par\vskip-8mm}
\def\bordero{						
	\setlength{\unitlength}{1mm}
	\newcount\xco
	\newcount\yco
	\xco=-24
	\yco=12
	\begin{picture}(140,0)
	\put(\xco,\yco){$\ktl$}
	\advance\yco by-1
	{\loop
	\put(\xco,\yco){$\kcl$}
	\advance\yco by-2
	\ifnum\yco>-240
	\repeat
	\put(\xco,\yco){$\kbl$}}
	\xco=158
	\yco=12
	\put(\xco,\yco){$\ktr$}
	\advance\yco by-1
	{\loop
	\put(\xco,\yco){$\kcr$}
	\advance\yco by-2
	\ifnum\yco>-240
	\repeat
	\put(\xco,\yco){$\kbr$}}
	\put(-20,12){\ooo
bacdefghidfghghdhededbihdgdfdfhhdheidhdhebaaahjhhdahbahgdedgehgfdiehhgdigicba}
	\put(-20,-241.5){\ooo
ababaighefdbfghgeahgdfgafagihdidihiidhiagfedhadbfdecdcdfagdcbhaddhbgfchbgfdacfediacbabab}
	\end{picture}
	\par\vskip-8mm}
\def\headpic{						
	\indent
	\setlength{\unitlength}{.4mm}
	\thinlines
	\par
	\begin{picture}(29,16)
	\put(165,16){\line(1,0){4}}
	\put(170,16){\line(1,0){4}}
	\put(180,16){\line(1,0){4}}
	\put(175,0){\line(1,0){4}}
	\put(180,0){\line(1,0){4}}
	\put(185,0){\line(1,0){4}}
	\put(169,0){\line(0,1){16}}
	\put(170,0){\line(0,1){16}}
	\put(179,0){\line(0,1){16}}
	\put(180,0){\line(0,1){16}}
	\put(184,0){\line(0,1){16}}
	\put(185,0){\line(0,1){16}}
	\put(169,16){\oval(8,32)[bl]}
	\put(170,16){\oval(8,32)[br]}
	\put(179,0){\oval(8,32)[tl]}
	\put(185,0){\oval(8,32)[tr]}
	\end{picture}
	\par\vskip-6.5mm
	\thicklines}
\def\title#1#2#3#4{\border\headpic {\hbox to\hsize{#4 \hfill UMDEPP #3}}\par
	\begin{center} \vglue .5in {\large\bf #1}\\[.6in]
	{#2}\\[.1in] {\it Department of Physics and Astronomy}\\
	{\it University of Maryland, College Park, MD 20742}\\[1.5in]
	{\bf Abstract}\\[.1in] \end{center} \begin{quotation}}	
\def\Title#1#2#3#4#5#6#7{\border\headpic
	{\hbox to\hsize{#7 \hfill UMDEPP #6}}\par
	\begin{center} \vglue .4in {\large\bf #1}\\[.4in]
	{#2}\\[.1in] {\it Department of Physics and Astronomy}\\
	{\it University of Maryland, College Park, MD 20742}\\[.1in]
	{#3}\\[.1in] {\it {#4}}\\ {\it {#5}}\\[.5in] {\bf Abstract}\\[.1in]
	\end{center} \begin{quotation}}			
\def\endtitle{\end{quotation}\newpage}			

\def\sect#1{\bigskip\medskip \goodbreak \noindent{\bf {#1}} \nobreak \medskip}
\def\refs{\sect{References} \footnotesize \frenchspacing \parskip=0pt}
\def\Item{\par\hang\textindent}

\def\doit#1#2{\ifcase#1\or#2\fi}
\def\[{\lfloor{\hskip 0.35pt}\!\!\!\lceil}
\def\]{\rfloor{\hskip 0.35pt}\!\!\!\rceil}

\def\du#1#2{_{#1}{}^{#2}}

\def\hati{{\hat{I}}}
\def\dt{$~D=10$~}

\def\pl#1#2#3{Phys.~Lett.~{\bf {#1}B} (19{#2}) #3}
\def\np#1#2#3{Nucl.~Phys.~{\bf B{#1}} (19{#2}) #3}

\def\pr#1#2#3{Phys.~Rev.~{\bf D{#1}} (19{#2}) #3}

\def\cmp#1#2#3{Comm.~Math.~Phys.~{\bf {#1}} (19{#2}) #3}

\def\ap#1#2#3{Ann.~of Phys.~{\bf {#1}} (19{#2}) #3}
\def\prep#1#2#3{Phys.~Rep.~{\bf {#1}C} (19{#2}) #3}
\def\ptp#1#2#3{Prog.~Theor.~Phys.~{\bf {#1}} (19{#2}) #3}
\def\ijmp#1#2#3{Int.~Jour.~Mod.~Phys.~{\bf {#1}} (19{#2}) #3}

\def\ibid#1#2#3{{\it ibid.}~{\bf {#1}} (19{#2}) #3}

\def\ula{{\un a}}
\def\ulb{{\un b}}
\def\ulc{{\un c}}
\def\uld{{\un d}}

\def\fracmm#1#2{{{#1}\over{#2}}}
\def\gg{{\hbox{\sc g}}}
\def\half{{\fracm12}}

\def\frac#1#2{{\textstyle{#1\over\vphantom2\smash{\raise -.20ex
	\hbox{$\scriptstyle{#2}$}}}}}			

\def\fracm#1#2{\hbox{\large{${\frac{{#1}}{{#2}}}$}}}

\def\Dot#1{\buildrel{_{_{\hskip 0.01in}\bullet}}\over{#1}}
\def\dt#1{\Dot{#1}}
\def\uln{{\underline n}}
\def\Tilde#1{{\widetilde{#1}}\hskip 0.015in}
\def\Hat#1{\widehat{#1}}

\def\scst{\scriptstyle}
\def\itrema{$\ddot{\scriptstyle 1}$}
\def\Bo{\bo{\hskip 0.03in}}
\def\lrad#1{ \left( A {\buildrel\leftrightarrow\over D}_{#1} B\right) }
\def\derx{\partial_x} \def\dery{\partial_y} \def\dert{\partial_t}
\def\Vec#1{{\overrightarrow{#1}}}
\def\.{.$\,$}

\def\grg#1#2#3{Gen.~Rel.~Grav.~{\bf{#1}} (19{#2}) {#3} }
\def\pla#1#2#3{Phys.~Lett.~{\bf A{#1}} (19{#2}) {#3}}

\def\ula{{\underline a}}
\def\ulb{{\underline b}}
\def\ulc{{\underline c}}
\def\uld{{\underline d}}
\def\ule{{\underline e}}
\def\ulf{{\underline f}}
\def\ulg{{\underline g}}
\def\ulm{{\underline m}}
\def\uln#1{\underline{#1}}
\def\ulp{{\underline p}}
\def\ulq{{\underline q}}
\def\ulr{{\underline r}}

\def\hatm{\hat m}
\def\hatn{\hat n}
\def\hatr{\hat r}
\def\hats{\hat s}
\def\hatt{\hat t}

\def\ul{\underline}
\def\un{\underline}
\def\-{{\hskip 1.5pt}\hbox{-}}

\def\kd#1#2{\d\du{#1}{#2}}
\def\fracmm#1#2{{{#1}\over{#2}}}
\def\footnotew#1{\footnote{\hsize=6.5in {#1}}}

\def\low#1{{\raise -3pt\hbox{${\hskip 1.0pt}\!_{#1}$}}}

\def\ip{{=\!\!\! \mid}}
\def\unb{{\underline {\bar n}}}
\def\upb{{\underline {\bar p}}}
\def\um{{\underline m}}
\def\up{{\underline p}}
\def\Phib{{\Bar \Phi}}
\def\Phit{{\tilde \Phi}}
\def\Phibt{{\tilde {\Bar \Phi}}}
\def\Db{{\Bar D}_{+}}
\def\gg{{\hbox{\sc g}}}
\def\nt{$~N=2$~}

\def\Dot#1{\buildrel{_{_{\hskip 0.01in}\bullet}}\over{#1}}
\def\dt#1{\Dot{#1}}
\def\gg{{\hbox{\sc g}}}
\def\nt{$~N=2$~}
\def\gg{{\hbox{\sc g}}}
\def\nt{$~N=2$~}
\def\tr{{\rm tr}}
\def\Tr{{\rm Tr}}
\def\mpl#1#2#3{Mod.~Phys.~Lett.~{\bf A{#1}} (19{#2}) #3}
\def\hati{{\hat i}} \def\hatj{{\hat j}} \def\hatk{{\hat k}}
\def\hatl{{\hat l}}

\begin{document}

\font\tenmib=cmmib10
\font\sevenmib=cmmib10 at 7pt 
\font\fivemib=cmmib10 at 5pt  
\font\tenbsy=cmbsy10
\font\sevenbsy=cmbsy10 at 7pt 
\font\fivebsy=cmbsy10 at 5pt  
\def\BMfont{\textfont0\tenbf \scriptfont0\sevenbf
                              \scriptscriptfont0\fivebf
            \textfont1\tenmib \scriptfont1\sevenmib
                               \scriptscriptfont1\fivemib
            \textfont2\tenbsy \scriptfont2\sevenbsy
                               \scriptscriptfont2\fivebsy}
\def\rlx{\relax\leavevmode}
\def\BM#1{\rlx\ifmmode\mathchoice
                      {\hbox{$\BMfont#1$}}
                      {\hbox{$\BMfont#1$}}
                      {\hbox{$\scriptstyle\BMfont#1$}}
                      {\hbox{$\scriptscriptstyle\BMfont#1$}}
                 \else{$\BMfont#1$}\fi}

\font\tenmib=cmmib10
\font\sevenmib=cmmib10 at 7pt 
\font\fivemib=cmmib10 at 5pt  
\font\tenbsy=cmbsy10
\font\sevenbsy=cmbsy10 at 7pt 
\font\fivebsy=cmbsy10 at 5pt  
\def\BMfont{\textfont0\tenbf \scriptfont0\sevenbf
                              \scriptscriptfont0\fivebf
            \textfont1\tenmib \scriptfont1\sevenmib
                               \scriptscriptfont1\fivemib
            \textfont2\tenbsy \scriptfont2\sevenbsy
                               \scriptscriptfont2\fivebsy}
\def\BM#1{\rlx\ifmmode\mathchoice
                      {\hbox{$\BMfont#1$}}
                      {\hbox{$\BMfont#1$}}
                      {\hbox{$\scriptstyle\BMfont#1$}}
                      {\hbox{$\scriptscriptstyle\BMfont#1$}}
                 \else{$\BMfont#1$}\fi}

\def\inbar{\vrule height1.5ex width.4pt depth0pt}
\def\sinbar{\vrule height1ex width.35pt depth0pt}
\def\ssinbar{\vrule height.7ex width.3pt depth0pt}
\font\cmss=cmss10
\font\cmsss=cmss10 at 7pt
\def\ZZ{\rlx\leavevmode
             \ifmmode\mathchoice
                    {\hbox{\cmss Z\kern-.4em Z}}
                    {\hbox{\cmss Z\kern-.4em Z}}
                    {\lower.9pt\hbox{\cmsss Z\kern-.36em Z}}
                    {\lower1.2pt\hbox{\cmsss Z\kern-.36em Z}}
               \else{\cmss Z\kern-.4em Z}\fi}
\def\Ik{\rlx{\rm I\kern-.18em k}}  
\def\IC{\rlx\leavevmode
             \ifmmode\mathchoice
                    {\hbox{\kern.33em\inbar\kern-.3em{\rm C}}}
                    {\hbox{\kern.33em\inbar\kern-.3em{\rm C}}}
                    {\hbox{\kern.28em\sinbar\kern-.25em{\rm C}}}
                    {\hbox{\kern.25em\ssinbar\kern-.22em{\rm C}}}
             \else{\hbox{\kern.3em\inbar\kern-.3em{\rm C}}}\fi}
\def\IP{\rlx{\rm I\kern-.18em P}}
\def\IR{\rlx{\rm I\kern-.18em R}}
\def\IN{\rlx{\rm I\kern-.20em N}}
\def\Ione{\rlx{\rm 1\kern-2.7pt l}}

\def\scst{\scriptstyle}
\def\itrema{$\ddot{\scriptstyle 1}$}
\def\Bo{\bo{\hskip 0.03in}}
\def\lrad#1{ \left( A {\buildrel\leftrightarrow\over D}_{#1} B\right) }
\def\derx{\partial_x} \def\dery{\partial_y} \def\dert{\partial_t}
\def\Vec#1{{\overrightarrow{#1}}}
\def\.{.$\,$}
\def\Check#1{\widecheck{#1}}

\def\grg#1#2#3{Gen.~Rel.~Grav.~{\bf{#1}} (19{#2}) {#3} }
\def\pla#1#2#3{Phys.~Lett.~{\bf A{#1}} (19{#2}) {#3}}

\def\ula{{\underline a}} \def\ulb{{\underline b}} \def\ulc{{\underline c}}
\def\uld{{\underline d}} \def\ule{{\underline e}} \def\ulf{{\underline f}}
\def\ulg{{\underline g}} \def\ulm{{\underline m}}
\def\uln#1{\underline{#1}}
\def\ulp{{\underline p}} \def\ulq{{\underline q}} \def\ulr{{\underline r}}

\def\hatm{\hat m}\def\hatn{\hat n}\def\hatr{\hat r}\def\hats{\hat s}
\def\hatt{\hat t}

\def\plpl{{+\!\!\!\!\!{\hskip 0.009in}{\raise -1.0pt\hbox{$_+$}}
{\hskip 0.0008in}}}
\def\mimi{{-\!\!\!\!\!{\hskip 0.009in}{\raise -1.0pt\hbox{$_-$}}
{\hskip 0.0008in}}}

\def\items#1{\\ \item{[#1]}}
\def\ul{\underline}
\def\un{\underline}
\def\-{{\hskip 1.5pt}\hbox{-}}

\def\kd#1#2{\d\du{#1}{#2}}
\def\fracmm#1#2{{{#1}\over{#2}}}
\def\footnotew#1{\footnote{\hsize=6.5in {#1}}}

\def\low#1{{\raise -3pt\hbox{${\hskip 1.0pt}\!_{#1}$}}}

\def\ip{{=\!\!\! \mid}}
\def\unb{{\underline {\bar n}}}
\def\upb{{\underline {\bar p}}}
\def\um{{\underline m}}
\def\up{{\underline p}}
\def\Phib{{\Bar \Phi}}
\def\Phit{{\tilde \Phi}}
\def\Phibt{{\tilde {\Bar \Phi}}}
\def\Db{{\Bar D}_{+}}
\def\gg{{\hbox{\sc g}}}
\def\nt{$~N=2$~}

\def\framing#1{\doit{#1}
{\framingfonts{#1}
\border\headpic
}}

\framing{0}

{}~~~
\doit0{\bf PRELIMINARY VERSION \hfill \today\\
}
\vskip 0.07in

{\hbox to\hsize{April 1994\hfill UMDEPP 94--127}}\par

\begin{center}

{\large\bf $~N=2$~ Superstring Theory Generates}\\
\vskip 0.01in
{\large\bf Supersymmetric Chern-Simons Theories}$\,$\footnote{This
work is supported in part by NSF grant \# PHY-91-19746.} \\[.1in]

\baselineskip 10pt

\vskip 0.25in

Hitoshi ~NISHINO\footnote{E-Mail (internet): Nishino@UMDHEP.umd.edu} \\[.2in]
{\it Department of Physics} \\ [.015in]
{\it University of Maryland at College Park}\\ [.015in]
{\it College Park, MD 20742-4111, USA} \\[.1in]
and\\[.1in]
{\it Department of Physics and Astronomy} \\[.015in]
{\it Howard University} \\[.015in]
{\it Washington, D.C. 20059, USA} \\[.18in]

\vskip 1.5in

{\bf Abstract}\\[.1in]
\end{center}

\begin{quotation}

{}~~~We show that the action of self-dual supersymmetric Yang-Mills theory
in four-dimensions, which describes the consistent massless background
fields for $~N=2$~ superstring, generates the actions for $~N=1$~ and $~N=2$~
supersymmetric non-Abelian Chern-Simons theories in three-dimensions after
appropriate dimensional reductions.  Since the latters play important roles for
supersymmetric integrable models, this result indicates
the fundamental significance of the $~N=2$~ superstring theory controlling
(possibly all) supersymmetric integrable models in lower-dimensions.

\endtitle

\def\Dot#1{\buildrel{_{{\hskip 0.01in}_\bullet}}\over{#1}}
\def\Dot#1{\raise-1.8pt\hbox{${_{_{_\bullet}}}\atop\hbox{$^{^{#1}}$}$}}

\def\dt#1{\Dot{#1}}
\def\gg{{\hbox{\sc g}}}
\def\nt{$~N=2$~}
\def\gg{{\hbox{\sc g}}}
\def\nt{$~N=2$~}
\def\tr{{\rm tr}}
\def\Tr{{\rm Tr}}
\def\mpl#1#2#3{Mod.~Phys.~Lett.~{\bf A{#1}} (19{#2}) #3}
\def\hati{{\hat i}} \def\hatj{{\hat j}} \def\hatk{{\hat k}}
\def\hatl{{\hat l}}

\oddsidemargin=0.03in
\evensidemargin=0.01in
\hsize=6.5in
\textwidth=6.5in

\centerline{\bf 1.~~Introduction}

The mathematical conjecture [1] that all the integrable models in
lower-dimensions would be generated by what is called self-dual Yang-Mills
(SDYM) theory [2] in four-dimensions ($D=4$)\footnotew{We sometimes
use also the notation $~D=(2,2)$~ in order to show the number of
positive and negative signs in the metric, respectively.  When such
signature does not matter, we simply use $~D=4$, instead.} has attracted much
attention in physics nowadays, due to the recent realization [3] that the
consistent massless background fields for $~N=2$~ open superstring theory are
nothing else than the SDYM fields or self-dual supersymmetric Yang-Mills
(SDSYM) fields.  In  particular, it has been discovered [4] that the
consistent background fields
for  $~N=2$~ open superstring must be $~N=4$~ SDSYM, while it must be $~N=8$~
self-dual supergravity (SDSG) for $~N=2$~ closed superstring, if they
are to be described by irreducible superfields.

Motivated by this development, we have recently studied various aspects of
SDSYM and self-dual supergravity systems with different number of
supersymmetries [5-8], which are obtained by some truncations of the maximally
supersymmetric SDSYM or SDSG theory above.

We have also shown that some well-known supersymmetric
integrable systems in lower-dimensions, such as supersymmetric KdV
equations [9], supersymmetric KP equations [10], supersymmetric
Wess-Zumino-Novikov-Witten models [11], topological
models and $~W_\infty$-~~~ gravity [12], as well as dilaton black-hole solution
[13] are indeed generated by the $~N=4$~ SDSYM theory after some dimensional
reductions (DRs).

Most of them strongly indicate that there must be
supersymmetric Chern-Simons (SCS) theory [14,15]
with the vanishing field strength in $~D=3$, serving as an intermediate theory
that connects the SDSYM theory in $~D=4$~ and these lower-dimensional
supersymmetric integrable models.  As a matter of fact, already in the context
of non-supersymmetric Chern-Simons (CS) theory there are lots of links
known between the CS theory and soluble lattice models, Yang-Baxter equations,
and monodromies of conformal field theories [16].  Therefore it is quite
natural
that the SCS theory has close relationship with the supersymmetric integrable
models or topological models.  Based on this indication, we try in this
Letter to show that the actions of the $~N=1$~ and $~N=2$~ SCS theories are
directly generated by the action of the SDSYM theory in $~D=4$~ after an
appropriate DR and truncation.  We show that this link is not just at the field
equation level, which has been already indicated in our previous  paper [15],
but is manifest at the action level after some peculiar DR.


\newpage

\centerline{\bf 2.~~CS Theory out of SDYM Theory}

Before we deal with the DR of the SDSYM theory, we try our DR for a purely
bosonic non-supersymmetric SDYM theory [2,17].

The suitable starting action for the SDYM theory, which is analogous to
the supersymmetric case later, is based on what is called Parkes-Siegel
action [4,17].  This action has a propagating fundamental multiplier
$~\Hat G^{\hat a\hat b\,I}$~ with the usual Yang-Mills field strength $~\Hat
F\du{\hat a\hat b}I$:
$$ I_{\rm SDYM} = -\half \int d^4 \Hat x \, \Hat G^{\hat
a\hat b \, I} \left(\Hat F_{\hat a \hat b}{}^I - \half\Hat\e\du{\hat a\hat
b}{\hat  c\hat d} \Hat F_{\hat c\hat d}{}^I \right) ~~.
\eqno(2.1)  $$
We use the universal notation that any fields or indices with {\it hats}
indicates those in $~D=4$, to be distinguished from the {\it non-hatted} fields
in $~D=3$~ treated later.  Therefore $~{\scst \hat a,\, \hat b,\,\cdots ~=~0,1,
2,3}$~ are for the world coordinates in $~D=4$, and relevantly we use the
signature $~(\Hat\eta_{\hat a\hat b}) = {\rm
diag.}\,(+,-,-,+)$.\footnotew{In this Letter we use the indices $~{\scst
0,1,2,3}$~ instead of $~{\scst 1,2,3,4}$, {\it unlike} our previous paper
[5].  The choice of the signature $~{\rm diag.}(+,-,-,+)$~ is for
convenience such that $~y\equiv x^3$~ can be the extra dimension.  In this
Letter, we do not use the underlined indices, in order to avoid
messy expressions.}  The
indices $~{\scst I,\,J,\,\cdots}$~ are for the adjoint representations
for the non-Abelian gauge group.  As is easily
seen, the field equation of $~\Hat G^{\hat a\hat b}$~
gives the self-duality of $~\Hat F_{\hat a\hat b}$, while it has some gauge
invariance under arbitrary shift in its self-dual part [4].
The lagrangian in (2.1) is by itself invariant under the Yang-Mills gauge
transformation, not to mention the total action.

We now try to setup some DR rule to get a CS theory action out of
(2.1).  First of all, we have to notice that the DR is not very
straightforward,
because the lagrangian of (2.1) {\it is} gauge invariant,
while the usual CS lagrangian [2] is {\it non-invariant}, yielding a total
divergence instead.  This indicates that the desirable DR we need should
break gauge at each field level, which is somehow recovered at
the action level in $~D=3$.  In other words, the gauge invariance of total
action is somehow restored under the specific DR prescription.  We
will clarify this point later.

Keeping this point in mind, we setup the following rule for the DR: First of
all, we regard the third coordinate $~\Hat x^3 \equiv y$~ to be periodic,
e.g., $~0 \le y < 2\pi$.  This requirement is not strict but advantageous,
when we perform $~y\-$integrations later.  We next setup the rule:
$$\li{ &\Hat A\du a I (\Hat x) = f(y) A\du a I (x) ~~, \cr
& \Hat A\du 3 I ( \Hat x) = 0 ~~,
&(2.2a) \cr
& \Hat G^{a b \, I } (\Hat x) = \e^{a b c} g(y) A\du c I (x) ~~,
\cr
& \Hat G^{a 3 \, I} (\Hat x) = - g(y) A^{a \, I} (x) ~~.
&(2.2b) \cr} $$
Under our universal rule, the indices $~{\scst a,~b,~\cdots ~=~0,~1,~2}$~ are
for
the  $~D=(1,2)$~ coordinates, and accordingly $~(x^a) =
(x^0,x^1,x^2)$~ with $~\e^{012} = +1$.  As is easily seen, (2.2b) is consistent
with the field  equation of the original $~\Hat G^{\hat a\hat b}$~ itself.
Out of (2.2a) we
can  construct the components of $~\Hat F$, as
$$\eqalign{ &\Hat F_{a b}{}^I(\Hat x) = f(y) \left[\, \partial_a A\du b I (x) -
\partial_b A\du a I(x) \, \right] + f^{I J K} f(y)^2 A\du a J (x) A\du b K
(x)~~,
\cr  & \Hat F_{a 3}{}^I (\Hat x) = - f'(y) A\du a I(x)~~,  \cr}
\eqno(2.3) $$
where $~f'(y) \equiv d f(y)/d y$.  Even though ~$\Hat F_{a b}
(\Hat x)$~ does not appear to be gauge covariant in $~D=3$, this will
eventually pose no problem, as will be discussed shortly.

Using (2.2) and (2.3) it is straightforward to approach our ``goal'' action
in $~D=3$~ in the form:
$$\eqalign{I_{\rm SDYM} = - \int d^3 x \int_0^{2\pi} d y \, \e^{a b c}
\Big[ g&(y) f(y) (2\partial_a A\du b I) A\du c I + g(y) f(y)^2 f^{I J
K} A\du a I A\du b J A\du c K  \cr
& - g(y) f'(y) A\du d I A^{d\, I} \, \Big] ~~.\cr}
\eqno(2.4) $$
{}From now on, we sometimes omit the $~x\-$dependence in $~A\du a I(x)$,
when it is clear from the environment.
As mentioned earlier, the $~y\-$integral here is periodic, and we can
specify the following integrals for the products of $~f(y) $~ and
$~g(y)$:
$$ \eqalign{ & \int_0^{2\pi}  d y\, g(y) f(y) = c~~, \cr
& \int_0^{2\pi} d y\, g(y) f(y)^2 = \fracm 2 3 c~~, \cr
& \int_0^{2\pi} d y\, g(y) f'(y) = 0~~, \cr }
\eqno(2.5) $$
where $~c$~ is some constant.  There are non-unique choices for $~f(y)$~
and $~g(y)$~ satisfying these conditions, but a natural and simple
example is
$$\eqalign{ & f(y) = \cos(my) ~~, \cr
&g(y) = \fracm 1 {3\pi} c \left[ 2 + 3 \cos(my) \right] ~~, ~~~~
(m = \pm 1,\pm 2, \cdots ) ~~, \cr }
\eqno(2.6) $$
where $~m$~ is an arbitrary non-zero integer, and the peculiar
cosine-functions are chosen such that
the periodic boundary conditions for $~\Hat A$~ and $~\Hat G$~ are
satisfied:
$$ \eqalign{&\Hat A\du {\hat a}I (x, y = 2\pi) =\Hat A\du {\hat a}I
(x, y =0) ~~, \cr
& \Hat G\du{\hat a \hat b} I (x, y = 2\pi) =\Hat G\du {\hat a \hat b}I
(x, y =0) ~~. \cr }
\eqno(2.7) $$
The coefficients of terms in (2.5) are fixed in such a way that the
final action $~I_{\rm CS}$~ has the right overall quantized coefficient
$~n/(16\pi)$~ [2], when we fix $~c$~ as
$$c = \fracmm n{16\pi} ~~, ~~~~ (n = \pm 1, \pm 2, \cdots) ~~,
\eqno(2.8) $$
for any non-Abelian gauge group whose $~\pi_3\-$homotopy mapping is
non-trivial.\footnotew{As some readers may have already noticed, the
number $~n$~ is related to the ``winding" number along the extra coordinate
$~y$~ in the integrals (2.5), namely if we have {\it e.g.,}~$~\int_0^{2\pi n}
d y\, g(y) f(y) = c n$, then the factor $~n$~ in (2.8) is automatically
generated for $~c = 1/(16\pi)$.}  After the
$~y\-$integration, (2.4) actually yields the $~D=3$~ CS  action, as
desired:
$$I_{\rm SDYM}~~{\buildrel{^{\rm DR}}\over\longrightarrow}~~I_{\rm CS} =
\fracmm
n {16\pi} \int d^3 x \, \e^{a b c} \left[ A\du a I F\du{b c} I - \fracm 1 3
f^{I
J K} A\du a I A\du  b J A\du c K \right]~~.
\eqno(2.9) $$

We may wonder about the loss of the original $~D=4$~ gauge covariance in
our DR rule (2.2).  We can understand this as a kind of ``hidden
symmetry'' of the original action, when the DR rule are restricted in a
peculiar way.  Intuitively, we can also understand that all the
apparently gauge non-covariant terms in $~\d \Hat F\du{\hat a\hat b}I $~
do not contribute to the variation of the total action in $~D=3$~ after
the $~y\-$integration.  We can see this in a more explicit computation, as
follows:  First, we rewrite $~\Hat F\du{a b} I $~ as
$$\Hat F\du{a b} I (\Hat x) = f(y) F\du{a b} I(x) + \left( f(y)^2 - f(y)\right)
f^{I J K} A\du a  J (x) A\du b K (x) ~~,
\eqno(2.10) $$
with manifest $~D=3$~ quantities.  Accordingly,
$~D=3$~ gauge transformations are
$$\eqalign{&\d \Hat F\du{a b} I (\Hat x) = - f(y) f^{I J K} \e^J (x)
F\du{a b} K (x) + 2 \left(f(y)^2 - f(y)\right) f^{I J K} A\du a J (x)
D_b\e^K (x) ~~, \cr
& \d \Hat F\du {a 3} I(\Hat x) = - f'(y) D_a \e^I (x) ~~, \cr
& \d \Hat G^{a b\, I} (\Hat x) = g(y) \e^{a b c} D_c \e^I (x) ~~ , \cr
& \d\Hat G^{a 3 \, I} (\Hat x) = - g(y) D^a \e^I (x) ~~. \cr }
\eqno(2.11) $$
The covariant derivative $~D_a$~ here is in terms of the usual $~D=3$~
gauge field $~A\du a I (x)$.  Applying these rules to (2.4), we get
$$\li{\d I_{\rm SDSYM} & = - \int d^3 x \int_0^{2\pi}  d y\, g(y) \e^{a b
c} f^{I J K} \left [ 3\left( f(y)^2 - f(y) \right) A\du a I A\du b J D_c \e^K
-f(y) F\du{a b} I A\du c J \e^K \right] \cr
& = c \int d^3 x \,\partial_a \left[ \e^{a b c} f^{I
J K} A\du b I A\du c J \e^K \right] ~~.
&(2.12)\cr}  $$
To get the last side we have arranged all the terms into two categories: (i)
Bilinear ~$A\-$terms  with one derivative, and (ii) Cubic $~A\-$terms.  For the
terms in (ii), we have used the Jacobi identity to show they all cancel each
other.  For terms in (i) we see their mutual cancellation after the
use of the $~y\-$integrals (2.5) up to a surface term as in the last side of
(2.12) under the $~x\-$integral, leaving the total action invariant.

The invariance of the total action after our ``non-covariant'' DR is, however,
very natural, since the resultant $D=3$~
lagrangian is exactly the CS lagrangian which leaves the action (but not the
lagrangian) invariant.  In other words, the advantage of the action level DR
is that once the CS action is obtained, it automatically guarantees the
validity of the non-covariant terms in the DR we have adopted.

\bigskip\bigskip

\centerline{\bf ~$~N=1$~ SCS out of $~N=1$~ SDSYM}

Once we have understood the DR rule for the non-supersymmetric case, it is
straightforward to generalize it to the supersymmetric case of SDSYM.
We start with the action for the $~N=4$~ SDSYM theory, which describes
the appropriate massless background fields of the $~N=2$~ superstring
theory [4].  As has been displayed in the papers [4,5], the maximally $~N=4$~
supersymmetric SDSYM theory [4] needs a peculiar multiplier superfield for its
lagrangian formulation based on what we call Parkes-Siegel formulation [4,17].
The field content of the $~N=4$~ SDSYM theory is  $~(\Hat A\du{\hat a}I, \Hat
G\du{\hat a\hat b} I, \Hat\r^{\,I},\Hat{\Bar\l}{}^I,  \Hat S\du i I, \Hat T\du
i
I)$,  where $~{\scst i,~j,~\cdots~=~1,~2,~3}$~ are the indices
for what are called $~\a$~ and $~\b\-$matrices acting on the $~N=4$~ indices
[5], and the {\it barred} fields\footnotew{This
convention is different from ref.~[5] in order to avoid using {\it tildes}
reserved for other purpose later.} are anti-chiral spinor fields.  Its
component
field action is given by [4,5]    $$ \li{ I\,_{\rm SDSYM}^{N=4} =
\,& \int d^4 x \Bigg[  - \fracm 12 \Hat G^{\hat a \hat b\,I}  (\Hat F\du{\hat a
\hat b}I- \half\Hat\e\du{\hat a \hat b}{\hat c \hat d}  \Hat F\du{\hat c \hat
d}
I) + \fracm 12  (\Hat D_{\hat a} \Hat S\du i I)^2 - \fracm 12 (\Hat D_{\hat a}
\Hat T\du i I)^2  + 2i (\Hat\r{}^I\Hat\G^{\hat a} \Hat D_{\hat a}
\Hat{\Bar\l}{}^I)  \cr  & ~~~~~ ~~~~~ - i f^{I J K} \left\{ (\Hat{\Bar\l}{}^I
\a_i \Hat{\Bar\l} {}^J) \Hat S\du i K + (\Hat{\Bar\l}{}^I \b_i\Hat{\Bar\l}{}^J)
\Hat T\du i K  \right\} \Bigg] ~~. &(3.1)  \cr } $$
where $~\Hat\G^{\hat a}$'s are gamma matrices in $~D=4$.

For our purpose of displaying a DR that gives a SCS theory, we can
simplify (3.1) into an action with lower supersymmetry by some
truncation.  A simple choice is to go down to $~N=1$~ SDSYM, namely we have the
(off-shell) prepotential superfield $~\Hat V$~ for $~(\Hat A\du{\hat\a}I,
\Hat{\Bar\l}\du{\hat{\Dot\a}}I; \Hat D^I, \Hat \l\du{\hat\a}I ) $~ and the
multiplier (off-shell) superfield $~\Hat\L^{\hat\a}$~ for $~(\Hat G\du{\hat
a\hat b}I, \Hat\r\du{\hat\a}I; \Hat\varphi^I,\Hat\psi\du{\hat\a} I)$~ with
the action [5]
$$\li{I\,_{\rm SDSYM}^{N=1} & = \int d^4 x \int d^2 \Hat\theta ~
\Hat  \L^{\hat\a \, I} \Hat W\du{\hat\a} I
&(3.2)  \cr
& = \int d^4 x \,\bigg[- \half \Hat G^{\hat a\hat b\, I}\left(
\Hat F\du{\hat a\hat b}I - \half\Hat\e\du{\hat a\hat b}{\hat c \hat d} \Hat
F\du{\hat c\hat d}I \right)   + i \Hat\r^{\,\hat\a\, I} (\Hat\G^{\hat
c})\du{\hat\a}{\hat{\Dot\b}} \Hat D_{\hat c}\Hat{\Bar\l}\du{\hat{\Dot\b}}I +
\Hat\varphi^I \Hat D^I + \Hat\psi^{\hat\a\, I} \Hat\l\du{\hat\a} I \bigg] ~~.
 \cr} $$
Due to its automatic and manifest closure
of supersymmetry, we work on superfield as much as possible from now on.
The multiplier superfield $~\Hat \L^{\hat\a}$~ is the superfield analog of the
$~\Hat G^{\hat a\hat b}\-$field in the previous non-supersymmetric case (2.1),
while  $~\Hat D^I,~\Hat\l^{\hat\a \, I}, ~\Hat\varphi^I~$ and
$~\Hat\psi\du{\hat\a} I$~ are auxiliary fields.  We can easily see in component
fields, which fields are to  be truncated in (3.1) to reach (3.2), but we skip
the details here.

We are now ready to setup our DR rule for (3.2) now in terms of superfields,
that will generate a SCS action in $~D=3$, which is a superspace
analog of the non-supersymmetric case (2.2):
$$\li{&\Hat A_{\hat A} (\Hat z) =
\cases{\Hat A_a (\Hat z) = f(y) A_a(z) ~~, \cr
 \Hat A_3(\Hat z) = f(y) A_3 (z) = 0 ~~, \cr
 \Hat A_{\hat\a} (\Hat z) = f(y) A_{\hat\a} (z)~~, \cr
\Hat {\Bar A}_{\hat{\Dot\a}} (\Hat z) = f(y) \Bar A_{\hat{\Dot\a}} (z) ~~,
\cr}
&(3.3a) \cr
& \Hat\L^{\hat\a}(\Hat z) = g(y) \L^{\hat \a} (z) ~~.
&(3.3b) \cr }$$
Here we are omitting the gauge indices $~{\scst I,~J,~\cdots}$~
temporarily, and $~(\Hat z^{\hat A}) = (\Hat x^{\hat a},\,
\Hat\theta^{\hat\a},\,\Hat{\Bar\theta}\raise5pt\hbox{$\scst\hat{\Dot\a}$})$~
are the $~D=4,\, N=1$~ superspace  coordinates, while $~(z^A) = (x^a,\,
\theta^\a)$~ are the \hbox{$~D=3,\,N=1$}  superspace coordinates.  Notice that
the numbers of the coordinates  $~\Hat{\theta}^{\hat\a}~{\scst (\hat\a ~=~
1,~2)}$~ and of  $~\theta^\a~{\scst (\a ~=~ 1,~2)}$~ are exactly the same, and
we can  even directly identify them, not withstanding a subtlety about the
difference in the inner products to be discussed later.  Considering this, we
can omit the {\it hat}-symbols on the  $~\theta\-$coordinates from now on.
Note that we can choose $~0\le y \equiv x^3 < 2\pi$~ as the previous
section, and the functions $~f$~ and $~g$~ are exactly the same as
before, satisfying (2.5).

We next express $~\Hat W^{\hat\a}$~ in terms of these superpotentials.  To this
end we use the notation in ref.~[5] with necessary {\it hats}.  After
some arrangement, we get
$$~\eqalign{ \Hat W^{\hat\a} (\Hat z) = &\, \fracmm i 2 (\Hat\G ^{\hat c}
) {}^{\hat\a\hat{\Dot\b}} \left[ \Hat{\Bar D}_{\hat{\Dot\b}} \Hat A_{\hat c}
(\Hat z) -
\Hat\partial_{\hat c} \Hat{\Bar A}_{\hat{\Dot\b}}  (\Hat z) + \[ \Hat{\Bar
A}_{\hat {\Dot  \b}} (\Hat z)  , \, \Hat A_{\hat c} (\Hat z)  \] \, \right] \cr
= & \, \fracmm i 2 (\Hat\G^{\hat c}){} ^{\hat\a\hat{\Dot\b}} \bigg[
f(y) F_{\hat{\Dot\b} c} (z) + (f(y)^2 - f(y) ) \[ {\Bar A}_{\hat{\Dot\b}} (z)
, \, A_c(z) \] \cr
& ~~~~~ ~~~~~ ~~~~~ + (\theta \Hat \G^y)_{\hat{\Dot\b}} f'(y) A_{\hat c}(z) -
\d_{\hat c \, 3} f'(y) {\Bar A}_{\hat{\Dot\b}} (z)  \bigg] ~~.  \cr }
\eqno(3.4) $$
If we plug this into the starting action (3.2), and perform the
$~y\-$integrals (2.5), we get
$$ \eqalign{ I\,{}_{\rm SDSYM}^{N=1} = & \, c \int d^3 x \int d^2 \theta\,
\Bigg[ A^{\hat\a}{}^I (z) \bigg\{ \fracmm i 2 (\Hat \G^c)\du {\hat\a} {\hat
{\Dot\b}} F_{\hat{\Dot\b} c}{}^I (z) \bigg\} \cr
& ~~~~~ ~~~~~ ~~~~~ ~~~~~
+ \fracmm i 6 (\Hat\G ^a)\du{\hat\b} {\hat{\Dot\g}} f^{I J K}
A^{\hat\b \, I}(z) \Bar A\du{\hat{\Dot\g}}J (z) A\du a K (z)\,\Bigg] ~~.\cr}
\eqno(3.5) $$
Here $~F_{\hat{\Dot\a}b} $~ is given purely by $~D=3$~ fields:
$$ F_{\hat{\Dot\a} b} (z) \equiv \Bar D_{\hat{\Dot\a}} A\du b I (z)  -
\partial_b \Bar A_{\hat{\Dot\a}} (z) + \[ \Bar A_{\hat{\Dot\a}} (z) ,
\, A_b(z) \]   ~~,
\eqno(3.6) $$
with the usage of the {\it hatted} $~D=4$~ spinorial indices
distinguished from the $~D=3$~ ones, as a reminder for the subtlety
about the inner products mentioned earlier.

We now have to deal with the DR of the fermionic indices.  For
this purpose, we first perform the DR for the gamma-matrices.  Using
the notation in ref.~[5], we get
$$\eqalign{ & \Hat \G^0 = \s_2 \otimes \t_3 = \s_2 \otimes \g^0 ~~, \cr
& \Hat \G^1 = \s_2 \otimes (-\t_1) = \s_2 \otimes\g^1 ~~, \cr
& \Hat\G^2 = \s_2 \otimes (-\t_2) = \s_2 \otimes \g^2 ~~, \cr
& \Hat \G^3 = \s_1 \otimes I_2 ~~. \cr }
\eqno(3.7) $$
Here $~\g^\m$~ can be identified with the gamma matrices in the final
$~D=(1,2)$~ [15], as is understood as follows:  According to the general
analysis
of fermions in diverse dimensions in  ref.~[18], a general Majorana spinor in
$~D=(1,2)$~ has the structure  $$\left( \Psi_\a \right) = \pmatrix{ \psi \cr
\psi ^*  \cr } ~~,  \eqno(3.8) $$
where the star-operation implies the complex conjugate.  Compared with
a two-component Majorana-Weyl spinor in $~D=(2,2)$~ [5], eq.~(3.8) has exactly
the same structure.  Therefore, we can directly identify the two
components in the latter with the former, and similarly for the $~\g\-$matrix
components in (3.7).  This direct DR rule for spinors works smoothly
everywhere,
{\it except for} a difference in the inner products of both cases, due to
different charge-conjugation matrices used as their ``metrics''.  To be more
specific, in the original $~D=(2,2)$, the  charge-conjugation matrix is defined
by [5]
$$\left(\Hat C_{\hat\a\hat\b} \right) = \s_3 \otimes \t_2 ~~,
\eqno(3.9) $$
while in the final $~D=(1,2)$, it is to be [18]
$$\left( C_{\a\b} \right) = i\t_2 ~~.
\eqno(3.10) $$
Due to the extra factor $~(+i)$~ in (3.10) compared with the chiral (upper two)
components in (3.9), we get an extra factor, when performing the replacement
for
the inner product
$$\psi^{\hat\a} \chi_{\hat \a} \longrightarrow (+i) \psi^\a \chi_\a
{}~~.
\eqno(3.11) $$
However, this extra factor $~(+i)$~ is exactly cancelled by
another factor ~$(-i)$~ coming out of the $~\s_2\-$matrix in (3.7), when
$~\Hat\G$'s are replaced by $~\g$'s.  Additionally, we can identify the
anti-chiral components $~\Bar A_{\hat{\Dot\a}}$~ in (3.3a) with $~A_\a$,
because
the anti-chiral spinors in $~D=(2,2)$~ are {\it independent} of each other,
instead of being complex conjugate to each other [5].

After these considerations in (3.5),
we see that
$$\eqalign{ I\,{}_{\rm SDSYM}^{N=1} ~~{\buildrel{^{\rm DR}}
\over\longrightarrow}~~ &
\fracmm n{16\pi}\int d^3 x  \int d^2 \theta \left[  A^\a(z) W_\a (z) - \fracmm
i
6 (\g^a)^{\b\g} A\du\b I (z) A\du \g J (z)  A\du a K (z) \, \right] \cr
&\,= I\,{}_{\rm SCS}^{N=1} ~~, \cr }
\eqno(3.12) $$
which is nothing else than the action of the $~N=1$~ SCS theory [15], because
$$\eqalign{ W_\a (z)
& = \fracmm i 2 (\g^c)\du \a\b F\du{\b c} I (z) \cr
& = \fracmm i 2 (\g^c)\du\a\b \left( D_\b A_c(z) - \partial_c
A_\b(z) + \[ A_\b (z) , \, A_c (z) \]  \right)\cr }
\eqno(3.13) $$
is the field strength superfield in $~D=(1,2),\,N=1$~
SCS theory [14,15]!  Thanks to the superfield notation, we do not have to worry
about the compatibility of our DR with supersymmetry.

The superficial gauge non-covariance of our DR rule (3.3) can
be again understood as the ``hidden'' symmetry of the original action
under our specific DR rule, as in the non-supersymmetric case,
whose detailed demonstration is skipped here.  We
have also seen that the analogy between the non-supersymmetric case and
the supersymmetric one is pretty parallel, even sharing the same functions
$~f(y)$~ and $~g(y)$.

\bigskip\bigskip\bigskip


\centerline {\bf 4.~~$N=2$~ SCS Theory out of $~N=2$~ SDSYM Theory}

Once we have understood the DR to get the $~N=1$~ SCS theory, it is
straightforward to think of higher supersymmetries, such
as $~N=2$~ SCS theory.  The parallel structure about fermions between
$~D=(2,2),\,N=2$~ and $~D=(1,2),\,N=2$~ also makes the DR easier, so that
we give the main result here.

The starting $~N=2$~ SDSYM action is in terms of two $~N=2$~ real superfields
$~\Hat\L^I$~ and $~\Hat S^I$~ [5]:
$$I\,{}_{\rm SDSYM}^{N=2} = \int d^4 x \int d^4 \theta~ \Hat\L^I \Hat S^I ~~.
\eqno(4.1) $$
This action is also easily obtained from the $~N=4$~ SDSYM action (3.1) for the
$~N=2$~ superstring, by appropriate truncations [5] similarly to the
previous $~N=1$~ case.

To appeal to the intuition of the readers to figure out the right DR rule, we
next give our ``goal'' action of the $~N=2$~ SCS theory [15]:
$$I\,{}_{\rm SCS}^{N=2} =
\fracmm n {16\pi} \int d^3 x \int d^4 \theta \int_0^1 d t \left[ \Tilde A\du tI
\left( \Tilde S^I + i k f^{I J K} \Tilde A^{\a \, J}  \Tilde{\Bar A}\du\a K
\right) + \hbox{c.c.} \right]~~.
\eqno(4.2) $$
In this section, the {\it barred} spinorial superfields in $~D=3$~ are Dirac
conjugate spinorial superfields [15].  The $~k$~ is an appropriate real
constant.  The
$~t\-$coordinate needed here is what we call ``Vainberg coordinate'' to make
the
lagrangian formulation possible in the $~N=2$~ SCS theory [15].  Accordingly,
all the {\it tilded} superfields $~\Tilde S, ~\Tilde A_\a$~ and ~$\Tilde {\Bar
A}_\a$~ satisfy the boundary conditions\footnotew{We are using the {\it checks}
instead of the {\it hats} in ref.~[15], in order not to be confused with the
hats for DRs.}
$$\eqalign{ &\Tilde S^I
(x,t=1,\zeta) = S^I(x,\zeta)  ~~, ~~~~ \Tilde S^I (x,t=0,\zeta) = 0~~, \cr
&\Tilde A_\a (x,t=1,\zeta) = A_\a(x,\zeta)~~, ~~~~ \Tilde A_\a (x,t=0,\zeta)
= 0 ~~, ~~~~~(\hbox{\it idem.~for~}\Tilde{\Bar A}_\a) ~~,  \cr}
\eqno(4.3) $$
where {\it non-tilded} superfields $~S^I(x,\zeta),~A_\a(x,\zeta)$~ and
$~\Bar A_\a(x,\zeta)$~ are purely $~D=3$~ superfields, and $~\zeta^\a$~
represents both $~\theta^\a$~ and $~\Bar\theta^\a$~ for the $~D=3,\,N=2$~
superspace.  The potential superfields are defined by  [15]
$$\eqalign{&\Tilde D_\a + \Tilde A_\a (x,t,\zeta) \equiv e^{\Tilde
V(x,t,\zeta)/2} \Tilde D_\a
e^{-\Tilde V(x,t,\zeta)/2} ~, ~~\Tilde{\Bar D}_\a + \Tilde{\Bar A}_\a
(x,t,\zeta) \equiv e^{- \Tilde V(x,t,\zeta)/2} \Tilde{\Bar D}_\a e^{\Tilde V
(x,t,\zeta)/2} ~, \cr
&\Tilde\partial_t + \Tilde A_t (x,t,\zeta) \equiv e^{\Tilde V(x,t,\zeta)/2}
\Tilde \partial_t e^{-\Tilde V(x,t,\zeta)/2}~, ~~\Tilde
{\Bar\partial}_t + \Tilde{\Bar A}_t (x,t,\zeta)
\equiv e^{-\Tilde V(x,t,\zeta)/2} \Tilde{\Bar\partial}_t e^{\Tilde V
(x,t,\zeta)/2} ~, \cr}
\eqno(4.4) $$

Our appropriate DR rule is now obvious due to the simple structure of the
final action (4.2).  To generate the desired form of (4.2), we first
identify the ``Vainberg coordinate'' $~t$~ in (4.2) with $~y\equiv x^3$~ in
$~D=4$, and then choose its range to be $~0 \le t \equiv y \equiv x^3 < 1$.
The DR rule can be established also by following the previous $~N=1$~ case,
in particular, the apparent similarity between the superfield contents in
$~D=3,\,N=2$~ SDSYM [5-8] and $D=3,\,~N=2$~ SCS [15], namely we
can identify the two sorts of $~\theta\-$coordinates in $~D=(2,2)$~ and
$~D=(1,2)$.  Thus the DR rule is eventually easier than before:
$$\eqalign{&\Hat\L^I(\Hat z) = \fracmm n{16\pi}\left[ \Tilde A\du t
I(x,t,\zeta) + \Tilde{\Bar A}\du t I (x,t,\zeta)  \right] ~~, \cr
& \Hat S^I (\Hat z) = \left[ \Tilde S^I (x,t,\zeta) + i k f^{I J K}
\Tilde A^{\a J}(x,t,\zeta) \Tilde{\Bar A}\du\a K (x,t,\zeta) \right] +
\hbox{c.c.}  \cr}
\eqno(4.5) $$

Applying this DR rule, we get the $~N=2$~ SCS theory [15] at one stroke,
because the integrand is simply
$$\eqalign{\Hat\L^I\Hat S^I & = \fracm
n{16\pi} \left[ (\Tilde A\du t I + \Tilde{\Bar A}\du t I) \Tilde S^I + (\Tilde
A\du t I + \Tilde{\Bar A}\du t I) i k \Big\{ \Tilde A^\a ,\Tilde{\Bar A}_\a
\Big\}{}^I \right] \cr
& = \fracm n{16\pi} \left[ \, \Tilde A\du t I \left( \Tilde S^I + i k \Big\{
\Tilde A^{\a \, I},\Tilde{\Bar A}_\a \Big\}{}^I \right) + \hbox{c.c.} \,\right]
\cr }
\eqno(4.6) $$
and using this in (4.1) yields
$$ I\,{}_{\rm SDSYM}^{N=2} ~~ {\buildrel{^{\rm DR}}\over\longrightarrow} ~~
I\,{}_{\rm SCS}^{N=2} ~~.
\eqno(4.7) $$
Compared with the previous $~N=1$~ case, the $~N=2$~ DR is much easier,
owing to the similarity already at the superfield level.

Interestingly enough, the ``Vainberg coordinate'' $~t$~ in the $~N=2$~ SCS
theory [15] exactly coincides with the ``extra coordinate'' $~x^3$~ in our DR
from  $~D=4$.  This
result strongly suggests the validity and naturalness of our DR rule to get SCS
theories out of the more fundamental $~D=4$~ SDSYM theory.  Relevantly,
the $~N=2$~ case needs no particular functions such as $~f(y)$~ or
$~g(y)$, compared with the previous $~N=1$~ case, because the ``Vainberg
coordinate'' is a built-in variable in the former.

\bigskip\bigskip\bigskip

\centerline {\bf 5.~~Concluding Remarks}

In this Letter, we have taken a very important step to show that
the actions for $~N=1$~ and $~N=2$~ SCS theories in $~D=(1,2)$~ are directly
generated by the $~N=1$~ and $~N=2$~ SDSYM theories in $~D=(2,2)$, which are
nothing else than truncated theories of the consistent $~N=4$~ SDSYM background
system for the $~N=2$~ open superstring theory.  As a by-product, we have also
shown that an analogous DR rule for the non-supersymmetric case works, as well.

To our knowledge, there has been no explicit demonstration that the $~D=4$~
SDYM action directly gives rise to the action of the CS theory, not to mention
the corresponding supersymmetric cases.  We  believe that our result has
provided the ``missing link'' between these two important theories, which
in turn indicates the importance of the $~N=2$~ (open) superstring theory.

In our prescription, we have used apparently gauge non-covariant DR rule,
which at first sight seemed meaningless.  However, we have seen that this can
be always understood as the realization of some ``hidden'' symmetry in the
original $~D=4$~ action, when the DR rule is specified and
the extra dimension is compactified, because all the gauge non-covariant
contributions in the variation of the action result only in a total
x-divergence
within ~$D=3$.

The success of our DR in the $~N=2$~ case is more appealing, because it
reveals the important aspect that the ``Vainberg coordinate'' exactly coincides
with the ``extra coordinate'' in the DR.  This strongly indicates
the validity and naturalness of our DR prescription related the topological
feature of the SCS theory.

Even though we did not demonstrate all the procedure in this Letter, it is more
straightforward to generate what is called BF-theory [19] or supersymmetric BF
theory from the original PS-formulation of SDYM theory or SDSYM theory.
This is because the original PS-formulation lagrangian (2.1) or (3.1) has
already a suggestive form of the lagrangian of the former theory.  However,
we stress the fact that not only BF-type theories, but also the more
non-trivial SCS theories themselves are directly generated by the SDSYM
theory in $~D=4$.

We have emphasized in this Letter the usage of actions instead of field
equations to demonstrate the mechanism of our DR, because some fields can be
truncated by hand in some DR for field equations, which even though is
straightforward at the classical level, becomes non-trivial at the quantum
level.  The action formulation is more explicit and manifest, when we have to
consider also the topological effects and gauge invariance, as well
as quantum effects in the system.  As a matter of fact, our DR rule that
appeared
to be gauge non-covariant, turned out to be free of problems, when considered
at
the action level.  Additionally, the quantization of the over-all coefficient
has the strong topological significance, only when formulated in terms of
action
principle.

There was a similar attempt [15,20] in the past in the context of $~N=1$~
heterotic superstring showing an interesting link between a CS theory and
$~D=10$~ background fields for the heterotic superstring.  Our result in the
present Letter indicates that $~N=2$~ superstring has even a closer
relationship
with SCS theories than the $~N=1$~ superstring.

We hope that our result has opened a new direction revealing the important link
between the $~N=2$~ superstring, SDSYM theories in $~D=4$~ and SCS theories
in $~D=3$, which in turn will play an important role for supersymmetric
topological as well as integrable models in lower-dimensions.

\bigskip\bigskip


We are indebted to S.J\.Gates, Jr.~and W\.Siegel for valuable suggestions.

\bigskip\bigskip

\vfill\eject

\refs

\def\\{\vskip 0.05in}
\def\item#1{\Item{#1}}
\def\items#1{\\ \item{[{#1}]}}

{\small

\items{1} M.F.~Atiyah, unpublished;
R.S\.Ward, Phil.~Trans.~Roy.~Lond.~{\bf
A315} (1985) 451;
N.J\.Hitchin, Proc.~Lond.~Math.~Soc.~{\bf 55} (1987) 59.

\items{2} A.~Schwarz, Lett.~Math.~Phys.~{\bf 2} (1978) 247;
W.~Siegel, \np{156}{79}{135};
J.~Schonfeld, \np{185}{81}{157};
R.~Jackiw and S.~Templeton, \pr{23}{81}{2291};
C.R.~Hagen, \ap{157}{84}{342}; \pr{31}{85}{331}.

\items{3} H.~Ooguri and C.~Vafa, \mpl{5}{90}{1389};
\np{361}{91}{469}; \ibid{367}{91}{83};
H.~Nishino and S.J.~Gates, Jr., \mpl{7}{92}{2543}.

\items{4} W.~Siegel, \pr{47}{93}{2504}.

\items{5} S.V.~Ketov, H.~Nishino and S.J.~Gates, Jr., \np{393}{93}{149};
S.V.~Ketov, S.J.~Gates, Jr.~and H.~Nishino, \pl{308}{93}{323}.

\items{6} H.~Nishino, S.J.~Gates, Jr. and S.V.~Ketov, \pl{307}{93}{331}.

\items{7} S.J.~Gates, Jr., H.~Nishino and S.V.~Ketov, \pl{297}{92}{99}.

\items{8} H.~Nishino, Maryland preprint, UMDEPP 93--79, to appear
in Int.~Jour.~Mod.~Phys.

\items{9} S.J.~Gates, Jr.~and H.~Nishino, \pl{299}{93}{255}.

\items{10} H.~Nishino, \pl{318}{93}{107}.

\items{11} H.~Nishino, \pl{316}{93}{293}.

\items{12} H.~Nishino, \pl{309}{93}{68}.

\items{13} H.~Nishino, Maryland preprint, UMDEPP 94--96, to appear in
Phys.~Lett.~B.

\items{14} S.J.~Gates, M.T.~Grisaru, M.~Ro{\v c}ek and W.~Siegel,
``{\it Superspace}'', (Benjamin/Cummings, Reading MA, 1983), page~27;
N.~Sakai and Y.~Tanii, \ptp{83}{90}{968}.

\items{15} H.~Nishino and S.J.~Gates, \ijmp{8}{93}{3371}.

\items{16} L.~Kauffman, Topology, {\bf 26} (1987) 395;
V.G.~Turaev, Inv.~Math.~{\bf 92} (1988) 527;
J.H.~Przytcki, Kobe J.~Math.~{\bf 4} (1988) 115;
E.~Witten, \cmp{121}{89}{351}.

\items{17} A.~Parkes, \pl{286}{92}{265}.

\items{18} T.~Kugo and P.K.~Townsend, \np{221}{83}{357}.

\items{19} G.~Horowitz, \cmp{125}{89}{417};
G.~Horowitz and M.~Srednicki, Comm.~Math.
Phys.~{\bf 130} (1990) 83;
H.~Blau, G.~Tompson, Ann.~Phys.~(NY), {\bf 205} (1991) 130;
D.~Birmingham, M.~Blau, M.~Rakowski, and G.~Tompson,
\prep{209}{91}{129}.

\items{20} H.~Nishino, \mpl{7}{92}{1805}.

}

\end{document}